\documentclass[twocolumn,iop, aps,superscriptaddress,showpacs]{revtex4-1}

\usepackage{mathptmx}
\usepackage{subfigure}
\usepackage{dcolumn}
\usepackage{amsmath,amssymb}
\usepackage{bm}
\usepackage{color}
\usepackage{latexsym}
\usepackage{epstopdf}
\usepackage{color}
\usepackage[english]{babel}
\usepackage{latexsym}

\usepackage{psfrag,graphicx} 
\usepackage{epsf} 
\usepackage{subfigure} 
\usepackage{amsmath} 
\usepackage{amssymb} 
\usepackage{amsfonts}
\usepackage{bm}
\usepackage{natbib}
\usepackage{epstopdf}
\usepackage{appendix}
\usepackage{changes}

\definecolor{mygrey}{gray}{0.35}
\definecolor{myblue}{rgb}{0.2,0.2,0.8}
\definecolor{myzard}{cmyk}{0,0,0.05,0}
\definecolor{mywhite}{rgb}{1,1,1}
\definecolor{myred}{rgb}{1,0.,0.3}

\usepackage[colorlinks=true,citecolor=myblue,linkcolor=myred]{hyperref}

\def\be{\begin{equation}}
\def\ee{\end{equation}}
\def\ba{\begin{align}}
\def\enda{\end{align}}
\def\bi{\begin{itemize}}
\def\ei{\end{itemize}}

 \def\ee{\mathord{\rm e}}

 \def\ee{\mathord{\rm e}}

\renewcommand{\ee}{{\rm e}}

\def\beq{\begin{equation}}
\def\beq{\begin{equation}}
\def\eeq{\end{equation}}

 \newcommand{\ket}[1]{|#1\rangle}

   \newcommand{\bla}[1]{\left(#1\right)}
 \newcommand{\blb}[1]{\left[#1\right]}

\begin{document}

\title[Short Title]{Refocusing two qubit gates with measurements for trapped ions}

\author{Tuvia Gefen}
\author{Daniel Cohen}
\author{Itsik Cohen}
\author{Alex Retzker}
\affiliation{Racah Institute of Physics, The Hebrew University of Jerusalem, Jerusalem 
91904, Givat Ram, Israel}
\date{\today}

\pacs{ 03.67.Ac,  03.67.-a, 37.10.Vz,75.10.Pq}

\begin{abstract}
{Dynamical decoupling techniques are the method of choice for increasing gate fidelities. While these methods have produced very impressive results in terms of decreasing local noise and increasing the fidelities of single qubit operations,  dealing with the noise of two qubit gates has proven more challenging. 
The main obstacle is that the noise time scale is shorter than the two qubit gate itself so that refocusing methods do not work.
We present a measurement and feedback based method to refocus two qubit gates which cannot be refocused by conventional methods.  We analyze in detail this method for an error model which is relevant for trapped ions quantum information.}
\end{abstract}
\maketitle

{\it Introduction ---}
Shor's discovery of an algorithm\cite{shor} to break the RSA cryptosystem demonstrated that  a quantum computer constitutes a promising system for tackling hard computational problems. However, it still remained unclear whether even conceptually a quantum computer can be constructed. One compelling reason is that even a minute error in each computational step would  rapidly accumulate  to a large error. Remarkably, the theory of quantum fault tolerance \cite{knill, aharonov,kitaev,Aliferis} showed that this intuition was wrong. Actually, in order for a quantum computer to output a correct result with and arbitrarily small probability of failure, each gate operation must only fail with a small probability below a certain threshold.  Thus, the precise value of the error threshold is extremely important to the field of quantum computation.  This has initiated an enormous effort to reduce gate infidelities below the fault tolerance threshold. Very good results have been achieved in various platforms, in particular in trapped ions, NV centers in diamond and superconducting devices. 



Dynamical decoupling methods have made it possible to exceed this threshold for single qubit gates \cite{ballance1,Du1}. However, in the case of two qubit gates, this task remains challenging. Most noise sources of the two qubit gates possess a sufficiently long correlation time. Therefore, dynamical decoupling has considerably improved two qubit gate fidelities in various platforms, and utilizing refocusing techniques are expected to improve these fidelities even more in the near future \cite{ballance1,Biercuk1,Timoney1,Tan,Webster,Timoney2}. However, the noise originating from laser or microwave amplitude fluctuations creates a noisy two body term that cannot be treated by dynamical decoupling, as it has a faster time scale than the gate duration.  This amplitude noise thus places a serious limitation on the fidelity of two qubit gates, and is believed to be a bottleneck that will impede future advances.

\begin{figure}[h]
\begin{center}
\includegraphics[width=0.45\textwidth]{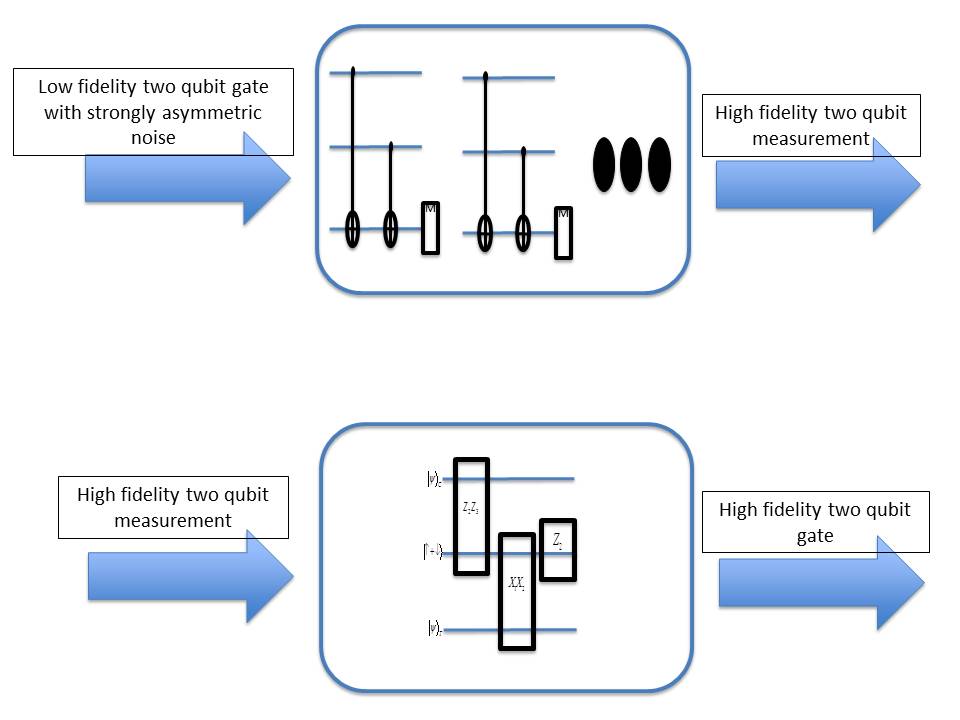}
\end{center}
\caption{The general scheme. The source of the protocol is a low fidelity two qubit gate with a specific error model with a single error operator. 
The first step is to utilize this two qubit gate together with single qubit operations to create efficient two qubit measurements, which are used in the next step to create high fidelity two qubit gates.}
\label{scheme}
\end{figure}

As a result most errors are below the fault tolerance threshold by more than an order of magnitude, but two qubit gates still suffer from a dominant amplitude error that is above it. In this paper, we show that in this scenario,  noisy two qubit gates can be refocused by a sequence of measurements and feedback, regardless of the noise's correlation time. The general idea is explained in fig. \ref{scheme}, and is based on the following reasoning. A faulty two qubit gate could be used to realize a faulty two qubit measurement.  The measurement however, could be repeated many times and reduce the infidelity substantially, resulting in a high fidelity two qubit measurement. In the next stage the high fidelity two qubit measurement could be utilized to create a high fidelity two qubit gate.  This general idea could be brought to use for many architectures having different error models, e.g., NV centers in diamond, and superconducting qubits. Yet, here we will concentrate mainly on trapped ion systems. The paper is structured as follows: the amplitude error in trapped ions is introduced, then we present our method and show that it can arbitrarily suppress this noise. We show a detailed fidelity analysis and discuss the relevance to other noise models.

{\it Entangling gates with trapped ions---}
Entangling gates with trapped ions are realized using the M{\o}lmer-S{\o}rensen (MS) scheme \cite{MS,Wineland2000N,Blatt2014NPH,Monroe2009PRL,Ozeri2014PRA, Tan2,Didi2003Nature,Lucas2015Nature1,Lucas2015arXiv}, which is constructed out of the following interaction:
\beq
H_{MS,i}=\Omega \bla{\sigma_{x,i}+\sigma_{x,A}} \bla{b^\dagger e^{-i\epsilon t} +b e^{i\epsilon t}},
\eeq
where $b^\dagger$, $(b)$ are the creation (annihilation) operators of the vibrational phonon, $\Omega$ is the sideband Rabi frequency, and $\epsilon=\omega_d-\nu$ is the detuning of the driving field from the secular frequency. During the MS gate operation, the two spins are entangled to the vibrational phonon, which generates the entanglement between the two spins. It is desirable to get a pure spin state; i.e., to keep the entanglement between the spins but to disentangle them from the phonons. This goal is indeed achieved, as after times of $\tau_{gate}=2\pi n/\epsilon$ with integer $n,$ the entanglement with the phonons is removed. In these times an effective Hamiltonian of $H=g\sigma_{X,1}\sigma_{X,2},$ where $g=\frac{\Omega^{2}}{\epsilon},$ is obtained.   
It is thus clear that in order to get a pure spin state, the accuracy of the gate duration, 
$\tau_{gate}$, must be high. This accuracy is determined by the stability of the detuning $\epsilon$. Taking advantage of the low drift in the trap frequency, and the high control of the driving frequency, the detuning $\epsilon$ remains stable during the experiment, and thus dynamical decoupling techniques \cite{Hayes2012PRA,Green2015PRL,Ivanov2015PRA} can be used to disentangle the phonon from the spins. 

Producing a pure spin state is also vulnerable to fluctuations of the Rabi frequency. These fluctuations may change the radius of the circle in the phonon phase space, such that even if the gate timing is accurate, there is a likelihood of not returning to the starting point; in other words, a phonon-spin entanglement may remain. Nevertheless, as long as these fluctuations are stable during a single circle, which is the case of the weak coupling regime (high detuning), the  phonon-spin entanglement is eliminated. In the strong coupling regime, this noise only makes a second order contribution, which is taken into account with the other noise terms that are not refocused.

However, the main source of decoherence originates from the first order contribution of the fluctuating Rabi frequency. These fluctuations eventually give rise to an amplitude noise in the interaction: instead of realizing an effective Hamiltonian of $g\sigma_{X,1}\sigma_{X,2}$ the following Hamiltonian: $(g+\Delta g)\sigma_{X,i}\sigma_{X,A}$ is realized. Note that this argument also holds for other entangling gate schemes with trapped ions, such as the two qubit phase gate \cite{Didi2003Nature,Lucas2015Nature1,Lucas2015arXiv}. In these two qubit gates, in addition to the noisy interaction, the fluctuating Rabi frequency gives rise to a single qubit noise, which can be refocused with regular dynamical decoupling techniques.

The noise in the interaction results in a faulty gate: 
\begin{eqnarray*}
U_{MS}=\exp \bla{-i \blb{\frac{\pi}{4}+\epsilon_i}\sigma_{x,1}\sigma_{x,2} }.
\end{eqnarray*}

This unitary is transformed to the CNOT gate using single-body operations: 

\begin{eqnarray*}
U_{Z,1} U_{x,2} U_{y,1} U_{MS} U_{y,1}^\dagger 
\end{eqnarray*}
with $U_{\alpha,k}=\exp\bla{-i\sigma_{\alpha,k}\pi/4}$ for $\alpha=x,y,z$ and $k=1,2$. Assuming perfect single body operations, the obtained gate is a perfect CNOT followed by an error of $\exp(i\epsilon \sigma_{Z,1}\sigma_{X,2}).$ The CNOT can be utilized to perform non local measurements, for example measurement of $\sigma_{Z,1}\sigma_{Z,2},$ i.e., parity detection, as shown in figure \ref{scheme}. The amplitude error will propagate through the CNOTs to a final $\exp(i(\epsilon_{1}\sigma_{Z,1}+\epsilon_{2}\sigma_{Z,2})\sigma_{X,3})
 $ error, which will result in an imperfect parity detection. This is illustrated in eq. \ref{parity_table}:   
\begin{eqnarray}
\ket{0}\ket{0}\ket{0}&\rightarrow& \cos(\epsilon_{1}+\epsilon_{2}) \ket{0}\ket{0}\ket{0} -i\sin\bla{\epsilon_1+\epsilon_2} \ket{0}\ket{0}\ket{1}\nonumber\\
\ket{0}\ket{1}\ket{0}&\rightarrow& \cos(-\epsilon_{1}+\epsilon_{2})\ket{0}\ket{1}\ket{1} +i\sin(-\epsilon_1 +\epsilon_2) \ket{0}\ket{1}\ket{0} \nonumber\\
\ket{1}\ket{0}\ket{0}&\rightarrow&  \cos(\epsilon_{1}-\epsilon_{2})\ket{1}\ket{0}\ket{1} +i\sin(\epsilon_1 -\epsilon_2) \ket{1}\ket{0}\ket{0}\nonumber\\
\ket{1}\ket{1}\ket{0}&\rightarrow&\cos(\epsilon_{1}+\epsilon_{2}) \ket{1}\ket{1}\ket{0} +i\sin(\epsilon_1+\epsilon_2) \ket{1}\ket{1}\ket{1}.
\label{parity_table}
\end{eqnarray}  
\begin{figure}[!t]
 	\centering
 	\includegraphics[scale=0.4]{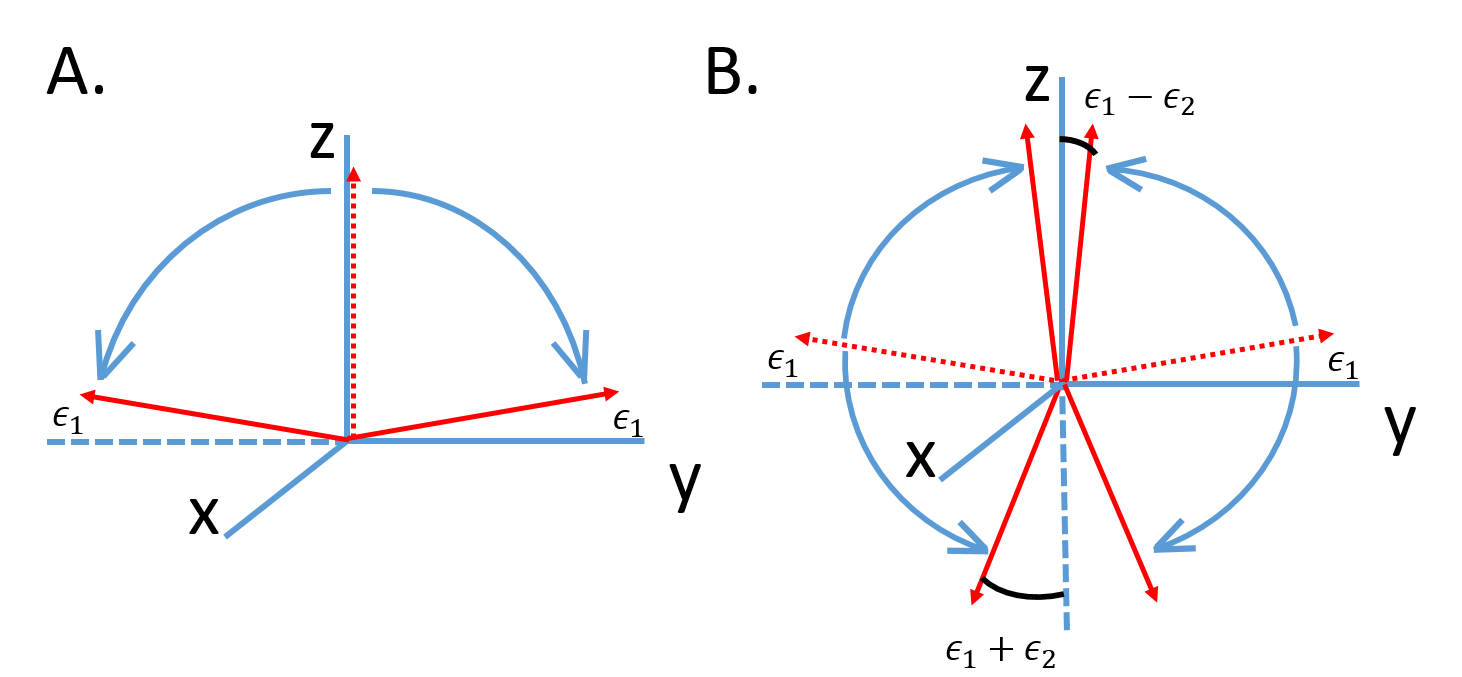}
 	\caption{(A) - The first imperfect CNOT gate rotates the state in $\pm(\frac{\pi}{4}+\epsilon_1)$ around the x axis. (B) - The second imperfect CNOT adds an additional rotation of $\pm(\frac{\pi}{4}+\epsilon_2)$ around the same axis depending on the second ions state. Hence, the errors in the CNOT's action can be the set ${\epsilon_1+\epsilon_2,\ -\epsilon_1+\epsilon_2,\ \epsilon_1-\epsilon_2, \ -\epsilon_1-\epsilon_2}$}   
 	\label{CNOT_rotation}
 \end{figure}
It can readily be observed that the error flips the ancilla, so that after the measurement there is still an overlap with states of the opposite parity. We will show that repeating the measurement many times reduces this overlap.

{\it The refocusing scheme--}  
{\it From low  fidelity CNOT to high fidelity measurement --- }
The basic advantage of a measurement over a gate is that the measurement's fidelity can be increased by repeating it a few times.  We now show the validity of this argument in our case. It can be seen that the noise does not only flip the ancilla, but also causes a dephasing. Fortunately, states with the same parity have the same deformation up to a constant relative phase that can be corrected; hence iteration is indeed useful. To illustrate, consider an initial state of $a|11\rangle+b|00\rangle+c|10\rangle+d|01\rangle.$ Detection of 0 , i.e. even parity, implies that we collapsed into: $\cos(e_{1}+e_{2})\left(a|11\rangle+b|00\rangle\right)+i\sin(e_{1}-e_{2})\left(c|10\rangle-d|01\rangle\right),$ and detection of 1 implies that we collapsed into: $i\sin(e_{1}+e_{2})\left(a|11\rangle-b|00\rangle\right)+\cos(e_{1}-e_{2})\left(c|10\rangle+d|01\rangle\right).$ As mentioned, there is an overlap with states of the opposite parity. If we now repeat the measurement three times and determine the parity according to a majority vote, the infidelity should go as $\sin(e)^{4}$ instead of $\sin(e)^{2}.$ This is verified by a simple examination of the trajectories: if the same outcome is obtained in all of the measurements, the overlap goes as $\sin(e)^{3},$ and the infidelity as $sin(e)^{6},$ but there are still trajectories in which not all the outcomes are the same. In these trajectories the overlap goes as $\sin(e),$ but the probability of these trajectories goes as $\sin(e)^{2}.$ Hence altogether this accounts for an infidelity that goes as $\sin(e)^{4}$ instead of $\sin(e)^{2}.$     
This can be easily generalized to $2n-1$ repetitions: the worst case scenario is an overlap that goes as $\sin(e),$ but the probability of these trajectories goes as $\sin(e)^{2n-2}$. Therefore performing $2n-1$ iterations reduces the infidelity to an order of magnitude of $e^{2n}.$ 

Note that this measurement scheme can be further improved to take fewer operations. First there is no need to apply two complete CNOT sequences in each iteration. It can be seen that $\sigma_{X,1}\sigma_{X,2}$ measurement is performed by simply applying $M_{Z,3}U_{MS,23}U_{MS,13},$ and therefore $\sigma_{Z,1}\sigma_{Z,2}$ measurement is realized by adding two Hadamards at the beginning of this sequence and two Hadamards at the end. Secondly, there is no need to perform the entire $2n-1$ repetitions in order to get a majority vote, we can end the sequence once any outcome is obtained $n$ times.
Further analysis of this scheme will be presented in the upcoming sections.             

{\it From high fidelity measurement to high fidelity CNOT--- }

\begin{figure}[h]
\begin{center}
\includegraphics[width=0.45\textwidth]{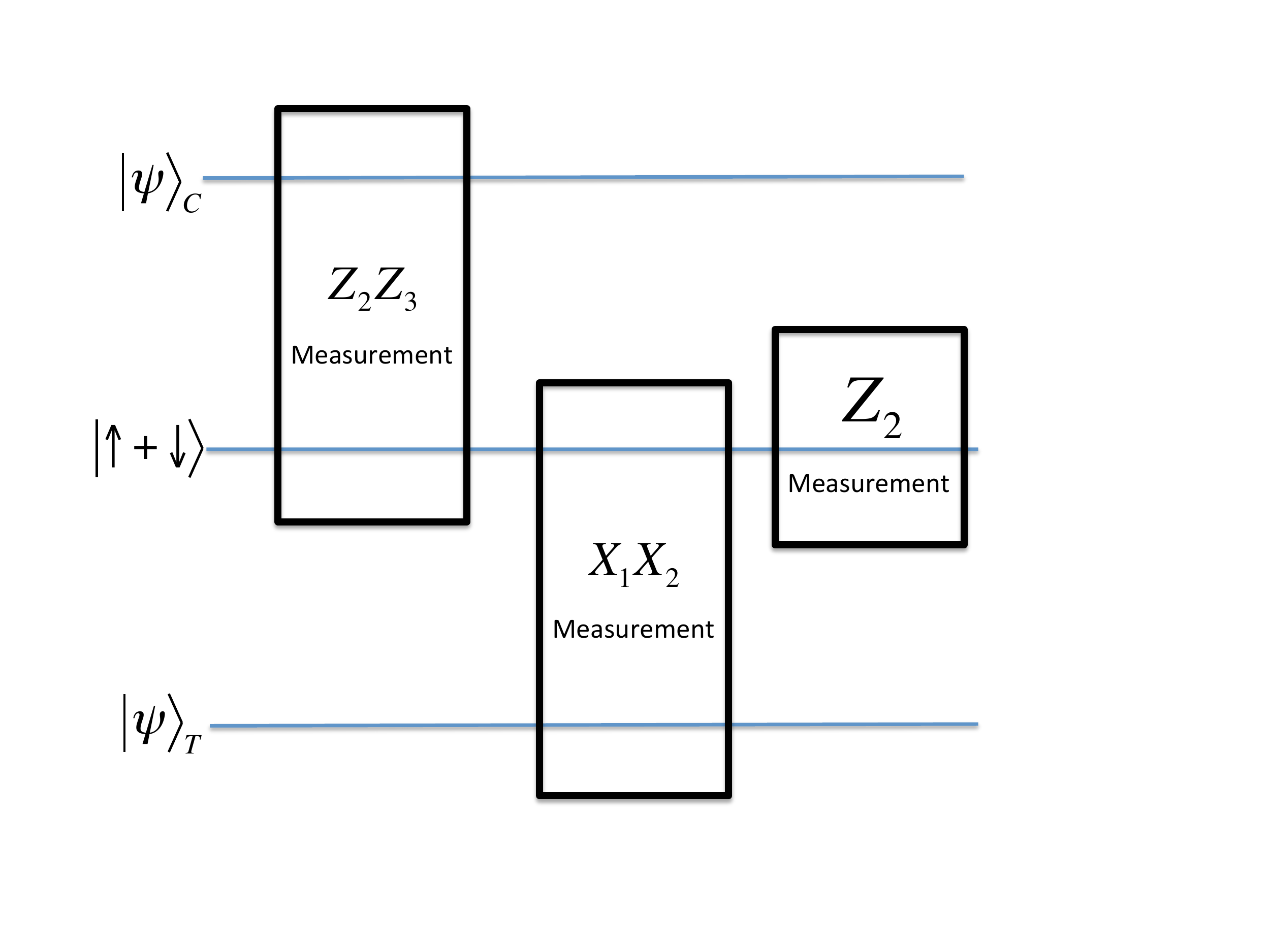}
\end{center}
\caption{ From two qubit measurement to a two qubit gate }
\label{CNOT_scheme}
\end{figure}

We now present our CNOT scheme which employs two body measurements and one body operations. This scheme is depicted in figure \ref{CNOT_scheme}, and is inspired by the scheme presented in \cite{gottesman}. Note that one ancillary qubit is required, as no measurement is performed solely on the original qubits and one more ancilla is needed to make the non-local measurements. The scheme is as follows:   
the initial state of the ancilla is set to $|0+1\rangle,$ so we start with  
\begin{equation}
\left(\alpha|11\rangle+\beta|00\rangle+\gamma|10\rangle+\delta|01\rangle\right)|0+1\rangle.
\end{equation}
Applying the measurement $\sigma_{Z,1}\sigma_{Z,3}$ we get:
\begin{equation}
\alpha|111\rangle+\beta|000\rangle+\gamma|101\rangle+\delta|010\rangle,
\end{equation}
otherwise a correction of $\sigma_{X,3}$ is applied.
In the next step we measure $\sigma_{X,2}\sigma_{X,3}$ and obtain:
\begin{equation}
\alpha|1(11+00)\rangle+\beta|0(00+11)\rangle+\gamma|1(01+10)\rangle+\delta|0(10+01)\rangle,
\end{equation}
otherwise a correction of $\sigma_{Z,1}\sigma_{Z,3}$ is applied.
And now by measuring the ancilla (third qubit) the CNOT is realized. 
 In this way a high fidelity CNOT gate can be realized under the condition that a high fidelity measurement can be realized. This specific example illustrates the main scenario in which there are high fidelity operations, i.e., the single qubit gates and measurements and one low fidelity operation: the two qubit CNOT. The fidelity of the single qubit operations is an order of magnitude below the threshold and the two qubit CNOT is an order of magnitude above the threshold. By utilizing many 'cheap' operations we are able to increase the fidelity of the bad operation and bring it below the threshold. 

We claim that this is the most efficient way to transform non-local measurements into a $CNOT$, since one cannot produce a CNOT using one two qubit measurement alone. This is quite clear: we can always choose a basis in which the operation of the measurement is not regular; i.e., two orthogonal states are mapped to the same state. Thus we cannot generate a CNOT utilizing only one two qubit measurement and single body operations.     

{\it Analysis ---} We first note that a parity measurement can be performed more efficiently. Instead of applying two CNOTs it is enough to realize the sequence: 
\begin{equation*}
M_{z,3} U_{y,2} U_{y,1} U_{MS2,3} U_{MS1,3} U_{y,2}^\dagger U_{y,1}^\dagger, 
\end{equation*}
where $M_{z,3}$ denotes a measurement of the ancilla. Similarly a measurement of $\sigma_{x,1}\sigma_{x,2}$ is realized by $M_{Z,3} U_{MS2,3} U_{MS1,3};$ in both cases a correction of single body operator is required according to the outcome. Due to the amplitude error the measurement is not accurate and there is an overlap with states of the opposite parity. So given an initial state of $a|11\rangle+b|00\rangle+c|10\rangle+d|01\rangle,$ if we detect an even parity we collapse into: $\cos(e_{1}+e2)\left(a|11\rangle+b|00\rangle\right)+\sin(e_{2}-e_{1})\left(c|10\rangle+d|01\rangle\right),$ and detection of an odd parity implies that we collapse into: $-\sin(e_{2}+e_{1})\left(a|11\rangle+b|00\rangle\right)+\cos(e_{2}-e_{1})\left(c|10\rangle+d|01\rangle\right).$ In order to decrease the infidelity, the measurement is repeated $2n-1$ times and the outcome is determined according to a majority vote. We now want to calculate the average infidelity, and show that if the infidelity of a single measurement goes as $e^{2}$, $2n-1$ repetitions yield infidelity that goes as
$e^{2n}.$ The average fidelity reads: 
\begin{equation}
\underset{\epsilon}{\sum}p(\epsilon)\underset{\psi}{\sum}\langle\psi(\epsilon)|\psi(\epsilon)\rangle|\langle\tilde{\psi}_{p}|\tilde{\psi}(\epsilon)\rangle|,
\end{equation}
where $\vert \psi(\epsilon) \rangle$ is the unnormalized wave-function given a specific measurement result, $\vert \tilde\psi(\epsilon) \rangle $ is the normalized one,
$|\tilde{\psi}_{p}\rangle$ is the normalized desired outcome. We observe that $|\tilde{\psi}_{p}\rangle=\frac{\Pi_{c}|\psi(\epsilon)\rangle}{\sqrt{\langle\psi(\epsilon)\Pi_{c}|\Pi_{c}\psi(\epsilon)\rangle}},$ where $\Pi_{c}$ is the projection on the correct subspace and $\Pi_{r}$
is the projection on the wrong subspace; thus the fidelity expression can be simplified to: 
\begin{equation}
\underset{\psi,\epsilon}{\sum}p(\epsilon)\langle\psi(\epsilon)|\psi(\epsilon)\rangle\sqrt{1-\frac{\langle\psi(\epsilon)\Pi_{r}|\Pi_{r}\psi(\epsilon)\rangle}{\langle\psi(\epsilon)|\psi(\epsilon)\rangle}}.
\label{fid2}
\end{equation}
Under the assumption that the probability of the odd subspace is of the same order of magnitude of the probability of the even subspace, which is indeed valid in all measurements performed in our CNOT, we get that: $\frac{\langle\psi(\epsilon)\Pi_{r}|\Pi_{r}\psi(\epsilon)\rangle}{\langle\psi(\epsilon)|\psi(\epsilon)\rangle}\sim e,$ and then taking the leading order of $e$ in eq. \ref{fid2} we get that
the infidelity is:
\begin{equation}
\frac{1}{2}\underset{\psi,\epsilon}{\sum}p(\epsilon)\langle\psi(\epsilon)\Pi_{r}|\Pi_{r}\psi(\epsilon)\rangle\approx\frac{1}{2}{n \choose \frac{n+1}{2}}\left(\frac{2e^{2}}{3}\right)^{\frac{n+1}{2}}.
\end{equation}  
This expression is correct only in the leading order of $e;$ a comparison with numerical values is shown in table \ref{tab}. 
\begin{table} 
\begin{tabular}{|c|c|c|}
\hline 
 & 3 repetitions & 5 repetitions\tabularnewline
\hline 
\hline 
Numerics & 0.0058 & 0.00137\tabularnewline
\hline 
Approximated & 0.0054 & 0.00108\tabularnewline
\hline 
\end{tabular}
\caption {Comparison of the approximated measurement infidelity and the values obtained by numerical integration.}
\label {tab}
\end{table}

This is obviously not enough, since we need to take into account other errors. We disregard preparation and detection errors as they are negligible and can be arbitrarily suppressed by iterations. This however is not true for single body errors caused by gates, which should be considered. Namely, each single body operation  is followed by a single body error with probability $\epsilon_{2},$ and each MS gate is followed by a single body error with probability $2\epsilon_{2}.$ We thus need to add the accumulation of these errors to the infidelity, when $\epsilon_{2}$ is assumed to be below the threshold.
 Note that we can reduce the number of MS gates, and thus reduce the accumulation of these errors because not all the repetitions are required. Instead of making all the $2n-1$ iterations, we just need to wait until one of the outcomes is repeated $n$ times. In the leading order of $e$ this yields the same infidelity as the $2n-1$ repetitions, but the average number of repetitions is reduced to $n(1+\frac{2e^{2}}{3})$ \cite{Supplamental Material}. The infidelity of the measurement thus reads: 
\begin{equation}
n\left(1+\frac{2e^{2}}{3}\right)4\epsilon_{2}+5\epsilon_{2}+\frac{1}{2}{2n-1 \choose n}\left(\frac{2e^{2}}{3}\right)^{n}.   
\end{equation}

 The infidelity of $\sigma_{X,i}\sigma_{X,j}$ is almost the same (it differs only in four Hadmard gates): $n(1+\frac{2e^{2}}{3})4\epsilon_{2}+\epsilon_{2}+\frac{1}{2}{2n-1 \choose n}(\frac{2e^{2}}{3})^{n}.$ We are now ready to obtain the infidelity of the CNOT. It can be seen that the amplitude error in the $\sigma_{Z,1}\sigma_{Z,3}$ measurement propagates to a final $\sigma_{X,2}$ error and the amplitude error in the $\sigma_{X,2}\sigma_{X,3}$ measurement propagates to a final $\sigma_{Z,1}$ error. Hence if our initial state is $\alpha|11\rangle+\beta|00\rangle+\gamma|10\rangle+\delta|01\rangle,$ the infidelity is:
\begin{eqnarray}
\begin{split}  
&\frac{1}{2}{2n-1 \choose n}\left(\frac{2e^{2}}{3}\right)^{n}\\
&\left[2-\left(|\beta|^{2}+|\delta|^{2}-|\alpha|^{2}-|\gamma|^{2}\right)-\left(\alpha*\delta+\delta*\alpha+\beta*\gamma+\gamma*\beta\right)\right]\\
&+8n\left(1+\frac{2e^{2}}{3}\right)\epsilon_{2}+9\epsilon_{2},
\end{split}
\end{eqnarray}   
which is bounded by: 
\begin{equation}
{2n-1 \choose n}\left(\frac{2e^{2}}{3}\right)^{n}+8n\left(1+\frac{2e^{2}}{3}\right)\epsilon_{2}+9\epsilon_{2}.
\end{equation}
 This bound is indeed attained for certain states.

\begin{figure}[t]
\vspace{+0.3cm}
\begin{center}
\includegraphics[width=0.5\textwidth]{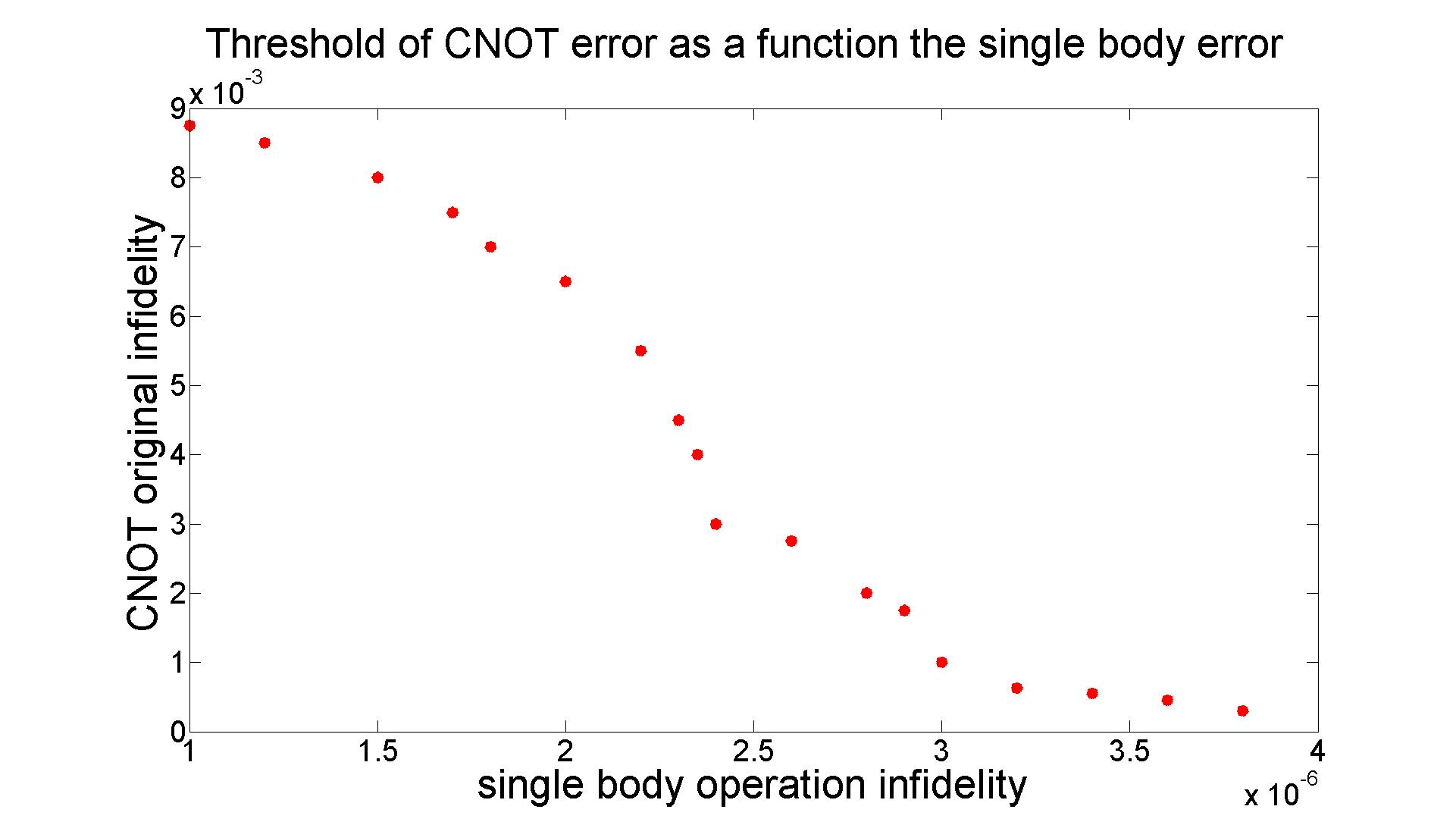}
\end{center}
\vspace{-0.3cm}
\caption{Given a threshold $T,$ the single body error $\epsilon_{2}$ determines a new threshold for the original CNOT error such that the infidelity of our CNOT will be below $T.$ The plot shows this new threshold as a function of $\epsilon_{2}.$ for $T=10^{-4}.$}
\label{new_threshold}
\end{figure}

  Therefore, given a threshold $T$, the single body error, $\epsilon_{2}$, limits the number of repetitions and dictates a new threshold for the original CNOT error which is lower than $T$. The threshold of the original CNOT as a function of $\epsilon_{2}$ is shown in figure \ref{new_threshold} for $T=10^{-4}.$  
  
{\it Relevance to NV centers and other noise models--} We have shown that our scheme works for amplitude error, which is highly important for trapped ions. This method can be applied to other noise models as well, in particular to dephasing in one of the qubits. Namely, instead of realizing a Hamiltonian of $g\sigma_{Z,1}\sigma_{Z,2},$ we get $g\sigma_{Z,1}\sigma_{Z,2}+\epsilon\sigma_{Z,1}.$ This noise again induces a syndrome mistake in the measurement and thus can be suppressed by iterating the measurement. 
This noise model is the dominant error in entangling gates in NV centers. In this platform the main source of decoherence is a slow drift in the NV energy gap, and this is the undesired $\epsilon\sigma_{Z,2}.$ When using these gates for parity measurements this noise flips the ancilla and changes the syndrome; thus the refocusing method works. \\
It should be noted that for other noise models such as dephasing in both qubits, our method is not useful, since the error does not only change the syndrome but also deforms the superposition. A future challenge would be to examine whether a similar method can be used to refocus this kind of noise.

{\it Conclusions and outlook---} We presented a new refocusing method that increases the fidelity of entangling gates, which suffer from dominant amplitude error or dephasing noise. The scheme is based on the observation that these gates can be used for accurate measurements that can function as building blocks for the desired entangling gate. Note that some noise models seem to be resilient to this method, which leaves us with the question of whether there are similar refocusing methods for them. Our scheme is based heavily on the fact that two body measurements performed by faulty entangling gates can be easily corrected by iterations; this however is not the case for other non-local multi-body measurements. Since these measurements are highly prevalent in error correction codes, it would be  interesting to inquire whether a similar correction method exists for them.     


\newpage
\newpage
\vspace{4cm}
\begin{widetext}
\section*{Supplemental Material}
\vspace{1cm}

{\it More elaborate analysis--}
Recall that the $\sigma_{Z,1}\sigma_{Z,2}$ measurement is performed in the following way: 
\begin{equation*}
M_{Z_{3}}\exp(i\frac{\pi}{4}\sigma_{Y,2})\exp(i\frac{\pi}{4}\sigma_{Y,1})\exp(i\frac{\pi}{4}\sigma_{X,2}\sigma_{X,3})\exp(i\frac{\pi}{4}\sigma_{X,1}\sigma_{X,3})\exp(-i\frac{\pi}{4}\sigma_{Y,2})\exp(-i\frac{\pi}{4}\sigma_{Y,1}),
\end{equation*}
 where $M_{z_{3}}$ denotes measurement of the third qubit (the ancilla),
and a correction of $\sigma_{Z,1}$ follows any measurement of $1$. Putting
the amplitude error: 
\begin{equation*}
M_{Z_{3}}\exp(i\frac{\pi}{4}\sigma_{Y,2})\exp(i\frac{\pi}{4}\sigma_{Y,1})\exp(i(\frac{\pi}{4}+e_{2})\sigma_{X,2}\sigma_{X,3})\exp(i(\frac{\pi}{4}+e_{1})\sigma_{X,1}\sigma_{X,3})\exp(-i\frac{\pi}{4}\sigma_{Y,2})\exp(-i\frac{\pi}{4}\sigma_{Y,1}),
\end{equation*}
 yields our imperfect measurement. The effect of the amplitude error is illustrated by the following table:
\begin{eqnarray*}
\begin{split} 
&|11\rangle|0\rangle\rightarrow i\cos(e_{1}+e_{2})|11\rangle|1\rangle-\sin(e_{1}+e_{2})|11\rangle|0\rangle\\		
&|10\rangle|0\rangle\rightarrow\cos(e_{1}-e_{2})|10\rangle|0\rangle+i\sin(e_{1}-e_{2})|10\rangle|1\rangle\\		
&|01\rangle|0\rangle\rightarrow\cos(e_{1}-e_{2})|01\rangle|0\rangle-i\sin(e_{1}-e_{2})|01\rangle|1\rangle\\		
&|00\rangle|0\rangle\rightarrow-i\cos(e_{1}+e_{2})|00\rangle|1\rangle+\sin(e_{1}+e_{2})|00\rangle|0\rangle.
\end{split}		 
\end{eqnarray*}		
 
  So given an initial state: $a|11\rangle+b|00\rangle+c|10\rangle+d|01\rangle,$
if we measure $1$ (and make the repair) we collapse into: 
\[
\cos(e_{1}+e2)\left(a|11\rangle+b|00\rangle\right)+\sin(e_{2}-e_{1})\left(c|10\rangle+d|01\rangle\right),
\]
and measuring $0,$ we collapse into: 
\[
-\sin(e_{2}+e_{1})\left(a|11\rangle+b|00\rangle\right)+\cos(e_{2}-e_{1})\left(c|10\rangle+d|01\rangle\right).
\]
Note that because of the amplitude error, the measurement is not accurate and there is an overlap with states of the opposite parity. This overlap goes as $\sin(e);$ thus, the infidelity of the measurement goes as $\sin(e)^{2}$. In order to decrease the infidelity, the measurement is repeated $2n-1$ times and the outcome is determined according to a majority vote. Let us consider the case of three repetitions: if we get the same outcome in all three measurements, the undesired overlap will go as $\sin(e)^{3}.$ For example, getting three times one will leave us with:
\begin{eqnarray}
\begin{split}  
&\cos(e_{1}+e2)\cos(e_{3}+e_{4})\cos(e_{5}+e_{6})\left(a|11\rangle+b|00\rangle\right)+\\
&\sin(e_{2}-e_{1})\sin(e_{4}-e_{3})\sin(e_{6}-e_{5})\left(c|10\rangle+d|01\rangle\right),
\end{split}
\end{eqnarray}
and thus the undesired overlap will be reduced. Regarding the case of different outcomes, for example two ones and
one zero, we obtain the following non-normalized state:
\begin{eqnarray}
\begin{split}
&-\cos(e_{1}+e2)\cos(e_{3}+e_{4})\sin(e_{5}+e_{6})\left(a|11\rangle+b|00\rangle\right)+\\
&\sin(e_{2}-e_{1})\sin(e_{4}-e_{3})\cos(e_{6}-e_{5})\left(c|10\rangle+d|01\rangle\right).
\end{split}
\end{eqnarray} 
After normalization the undesired overlap goes as $\sin(e),$ but the probability for this trajectory goes as $\sin(e);$ altogether this accounts for an infidelity that goes as $\sin(e)^{4}$, instead of $\sin(e)^{2}$.   
 
We would now like to make a more precise calculation of the average fidelity of this measurement, given a certain number of repetitions. We wish
to show that given a single measurement with an infidelity that goes
as $e^{2}$, then $2n-1$ repetitions yield infidelity that goes as
$e^{2n}$. The average fidelity reads: 
\[
\underset{\epsilon}{\sum}p(\epsilon)\underset{\psi}{\sum}\langle\psi(\epsilon)|\psi(\epsilon)\rangle|\langle\tilde{\psi}_{p}|\tilde{\psi}(\epsilon)\rangle|,
\]
 where $\vert \psi(\epsilon) \rangle$ is the unnormalized wave function given a specific measurement result, $\vert \tilde\psi(\epsilon) \rangle $ is the normalized one,
  $|\tilde{\psi}_{p}\rangle$ is the normalized desired outcome,
we observe that $|\tilde{\psi}_{p}\rangle=\frac{\Pi_{c}|\psi(\epsilon)\rangle}{\sqrt{\langle\psi(\epsilon)\Pi_{c}|\Pi_{c}\psi(\epsilon)\rangle}},$
where $\Pi_{c}$ is the projection on the correct subspace and $\Pi_{r}$
is the projection on the wrong subspace. Hence the average fidelity reads:
\begin{equation}
\underset{\psi,\epsilon}{\sum}p(\epsilon)\sqrt{\langle\psi(\epsilon)\Pi_{c}|\Pi_{c}\psi(\epsilon)\rangle\langle\psi(\epsilon)|\psi(\epsilon)\rangle}.
\label{fid1}
\end{equation}
 The average fidelity is thus larger than $\underset{\psi,\epsilon}{\sum}p(\epsilon)\langle\psi(\epsilon)\Pi_{c}|\Pi_{c}\psi(\epsilon)\rangle,$
and therefore the infidelity is bounded by $\underset{\psi,\epsilon}{\sum}p(\epsilon)\langle\psi(\epsilon)\Pi_{r}|\Pi_{r}\psi(\epsilon)\rangle.$
This bound is the same for any initial state, and in the case of three
repetitions we have: $\underset{\psi,\epsilon}{\sum}p(\epsilon)\langle\psi(\epsilon)\Pi_{r}|\Pi_{r}\psi(\epsilon)\rangle=S(e)^{3}+{3 \choose 2}C(e)S(e)^{2},$
where $C(e)=\frac{1}{4 e^2}\int_{-e}^e  \int_{-e}^e\cos^{2}(x+y)dxdy\sim1-\frac{2e^{2}}{3}$
and $S(e)=\frac{1}{4 e^2}\int_{-e}^e \int_{-e}^e\sin^{2}(x+y)dxdy\sim\frac{2e^{2}}{3}.$
Hence for $3$ repetitions we get that the infidelity is bounded by
$3\left(\frac{2e^{2}}{3}\right)^{2},$ and for $n$ repetitions it
is bounded by ${n \choose \frac{n+1}{2}}\left(\frac{2e^{2}}{3}\right)^{\frac{n+1}{2}}.$
This bound is attained for any initial state that lies in only one
of the subspaces (any initial state with a well defined parity). This is
because in this case for any trajectory that gives the correct parity
$\Pi_{c}|\psi(\epsilon)\rangle=|\psi(\epsilon)\rangle,\Pi_{r}|\psi(\epsilon)\rangle=0$
and for any trajectory that gives the wrong parity $\Pi_{r}|\psi(\epsilon)\rangle=|\psi(\epsilon)\rangle,\Pi_{c}|\psi(\epsilon)\rangle=0.$
However in the case where $|a|^{2}+|b|^{2}$ is of the same order of magnitude
of $|c|^{2}+|d|^{2},$ the infidelity will be smaller than the bound.
We now calculate the infidelity in this case: note that eq.\ref{fid1} can
be written as: 
\[
\underset{\psi,\epsilon}{\sum}p(\epsilon)\sqrt{\langle\psi(\epsilon)|\psi(\epsilon)\rangle^{2}-\langle\psi(\epsilon)\Pi_{r}|\Pi_{r}\psi(\epsilon)\rangle\langle\psi(\epsilon)|\psi(\epsilon)\rangle},
\]
that can be simplified to: 
\begin{equation}
\underset{\psi,\epsilon}{\sum}p(\epsilon)\langle\psi(\epsilon)|\psi(\epsilon)\rangle\sqrt{1-\frac{\langle\psi(\epsilon)\Pi_{r}|\Pi_{r}\psi(\epsilon)\rangle}{\langle\psi(\epsilon)|\psi(\epsilon)\rangle}}.
\label{fid2_supp}
\end{equation}
Under the assumption that $|a|^{2}+|b|^{2}$ has the same order of
magnitude of $|c|^{2}+|d|^{2},$ then $\frac{\langle\psi(\epsilon)\Pi_{r}|\Pi_{r}\psi(\epsilon)\rangle}{\langle\psi(\epsilon)|\psi(\epsilon)\rangle}\sim e$
and then taking the leading order of $e$ in eq. \ref{fid2_supp} we get that
the infidelity is: 
\begin{equation}
\frac{1}{2}\underset{\psi,\epsilon}{\sum}p(\epsilon)\langle\psi(\epsilon)\Pi_{r}|\Pi_{r}\psi(\epsilon)\rangle\approx\frac{1}{2}{n \choose \frac{n+1}{2}}\left(\frac{2e^{2}}{3}\right)^{\frac{n+1}{2}}.
\end{equation}
 We can compare this approximation to numerical values for $e=0.3,$
and $|a|^{2}+|b|^{2}=|c|^{2}+|d|^{2}$ : %
\begin{tabular}{|c|c|c|}
\hline 
 & 3 repetitions & 5 repetitions\tabularnewline
\hline 
\hline 
numerical & 0.0058 & 0.00137\tabularnewline
\hline 
Approximated & 0.0054 & 0.00108\tabularnewline
\hline 
\end{tabular} .

We also need to take into account other single body errors that are assumed to be below the threshold. As mentioned in the main part, we neglect preparation and detection errors as they can be arbitrarily suppressed by iteration. We do not neglect errors in gates and assume the following noise model: each single body rotation of $\frac{\pi}{4}$ is followed by a single body error with probability $\epsilon_{2}$ and any MS gate is followed by a single body error with probability $\epsilon_{2}$ on each of the relevant two qubits. Now $2n-1$ iterations of $\sigma_{Z,i}\sigma_{Z,j}$ include $4n-2$ MS gates, 4 Hadmard gates and one correction of $\frac{\pi}{2}$ rotations with probability $\frac{1}{2}.$ We thus get that the infidelity of this measurement is $(8n+1)\epsilon_{2}+\frac{1}{2}{2n-1 \choose n}\left(\frac{2e^{2}}{3}\right)^{n}.$ But, as mentioned in the main text, the number of single body errors can be significantly reduced, as we do not need all the $2n-1$ repetitions. Recall that $2n-1$ repetitions are used to lower the amplitude error from $\epsilon$ to $\epsilon^{n}.$ But in order to get this improvement we just need to wait until one of the outcomes is repeated $n$ times, and in most cases this will require less than $2n-1$ repetitions. So a more efficient scheme is to repeat the measurement until one of the outcomes occurs $n$ times; in the leading order of $e$ the same infidelity is achieved (taking into account only the amplitude errors). Note that the probability to stop after $k$ repetitions ($n\leq k \leq 2n-1$) is ${n+k-1 \choose k}\left(C(e)^{n}S(e)^{k}+S(e)^{n}C(e)^{k}\right),$ thus the average stopping time is given by $\sum_{k=0}^{n-1}k{n+k-1 \choose k}\left(p^{n}(1-p)^{k}+p^{k}(1-p)^{n}\right),$ which goes as $n(1+S(e)).$ Hence, the probability for a single body error is now $[4n(1+S(e))+5]\epsilon_{2},$ and the infidelity reads:
\begin{equation}
\left[4n(1+\frac{2e^{2}}{3})+5\right]\epsilon_{2}+\frac{1}{2}{2n-1 \choose n}\left(\frac{2e^{2}}{3}\right)^{n}. 
\end{equation}

We now want to calculate the average infidelity of the entire CNOT.
The imperfect $\sigma_{X,1}\sigma_{X,2}$ measurement yields the following outcomes:\\
 Collapsing into $|0\rangle$ (so we do not need to perform a correction):
\begin{eqnarray*}
|11\rangle\rightarrow\cos(e_{1}-e_{2})|11-00\rangle-\sin(e_{1}+e_{2})|11+00\rangle\\
|10\rangle\rightarrow\cos(e1-e2)|10-01\rangle-\sin(e1+e2)|10+01\rangle\\
|01\rangle\rightarrow\cos(e1-e2)|01-10\rangle-\sin(e1+e2)|01+10\rangle\\
|00\rangle\rightarrow\cos(e1-e2)|00-11\rangle-\sin(e1+e2)|00+11\rangle
\end{eqnarray*}
 Collapsing into $|1\rangle$ (and applying $\sigma_{X,1}$ correction):
\begin{eqnarray*}
|11\rangle\rightarrow\cos(e1+e2)|11+00\rangle+\sin(e2-e1)||11-00\rangle\\
|10\rangle\rightarrow\cos(e1+e2)|10+01\rangle+\sin(e2-e1)|10-01\rangle\\
|01\rangle\rightarrow\cos(e1+e2)|01+10\rangle+\sin(e2-e1)|01-10\rangle\\
|00\rangle\rightarrow\cos(e1+e2)|00+11\rangle+\sin(e2-e1)|00-11\rangle &  & .
\end{eqnarray*}
The infidelity analysis of this measurement is obviously similar to
that of the $\sigma_{Z,1}\sigma_{Z,2}$ measurement. except that one does not need the four Hadamard gates, so the infidelity of this measurement is: $\left[4n(1+\frac{2e^{2}}{3})+1\right]\epsilon_{2}+\frac{1}{2}{2n-1 \choose n}\left(\frac{2e^{2}}{3}\right)^{n}$. \\
The CNOT consists of $\sigma_{Z}\sigma_{Z}$ and $\sigma_{X}\sigma_{X}$ measurements, which we assume
are the only source of errors. Then, in order to calculate the infidelity
we just need to understand how these errors propagate in the CNOT.
Let us examine this for the $\sigma_{Z,1}\sigma_{Z,2}$ measurement:
\begin{equation}
\left(\alpha|11\rangle+\beta|00\rangle+\gamma|10\rangle+\delta|01\rangle\right)|1+0\rangle,
\end{equation}
 will become after the measurement: 
\begin{eqnarray*}
\begin{split}
&r(\epsilon)\left(\alpha|111\rangle+\beta|000\rangle+\gamma|101\rangle+\delta|010\rangle\right)+\\
&\epsilon\left(\alpha|110\rangle+\beta|001\rangle+\gamma|100\rangle+\delta|011\rangle\right),
\end{split}
\end{eqnarray*}
 where $r(\epsilon)$ denotes the relevant normalization factor. After a flawless $\sigma_{X,2}\sigma_{X,3}$ measurement (and repair if necessary)
and a $\sigma_{Z,2}$ measurement (and repair if necessary) we get: 
\begin{equation}
r(\epsilon)(\alpha|10\rangle+\beta|00\rangle+\gamma|11\rangle+\delta|01\rangle)+\epsilon(\alpha|11\rangle+\beta|01\rangle+\gamma|10\rangle+\delta|00\rangle).
\end{equation}
 Hence $\sigma_{Z,1}\sigma_{Z,2}$ error corresponds to an $\sigma_{X,2}$ error. \\
Considering the $\sigma_{X,2}\sigma_{X,3}$ error: 
\begin{equation}
\alpha|111\rangle+\beta|000\rangle+\gamma|101\rangle+\delta|010\rangle,
\end{equation}
will become after the measurement: 
\begin{eqnarray*}
\begin{split}
&r(\epsilon)\left(\alpha|1(11+00)\rangle+\beta|0(00+11)\rangle+\gamma|1(10+01)\rangle+\delta|0(01+10)\rangle\right)+
\\
&\epsilon\left(\alpha|1(11-00)\rangle+\beta|0(00-11)\rangle+\gamma|1(01-10)\rangle+\delta|0(10-01)\rangle\right).
\end{split}
\end{eqnarray*}
And after measurement of $\sigma_{Z,3}:$ 
\begin{equation}
r(\epsilon)\left(\alpha|10\rangle+\beta|00\rangle+\gamma|11\rangle+\delta|01\rangle\right)+\epsilon\left(-\alpha|10\rangle+\beta|00\rangle-\gamma|11\rangle+\delta|01\rangle\right).
\end{equation}
Hence the $\sigma_{X,2}\sigma_{X,3}$ error corresponds to a $\sigma_{Z,1}$ error. So
after neglecting the terms that correspond to error in both measurements,
we get: 
\begin{eqnarray*}
\begin{split}
&r(\epsilon_{1},\epsilon_{2})\left(\alpha|10\rangle+\beta|00\rangle+\gamma|11\rangle+\delta|01\rangle\right)+\epsilon_{1}\left(\alpha|11\rangle+\beta|01\rangle+\gamma|10\rangle+\delta|00\rangle\right)+
\\
&\epsilon_{2}\left(-\alpha|10\rangle+\beta|00\rangle-\gamma|11\rangle+\delta|01\rangle\right).
\end{split}
\end{eqnarray*}
In the leading order of $e,$ the infidelity equals $\frac{1}{2}\underset{\psi,\epsilon}{\sum}p(\epsilon_{1},\epsilon_{2})\langle\psi(\epsilon_{1},\epsilon_{2})\Pi_{r}|\Pi_{r}\psi(\epsilon_{1},\epsilon_{2})\rangle.$
So this is just: 
\begin{eqnarray*}
\underset{\epsilon}{\sum}p(\epsilon_{1},\epsilon_{2})\left[(\epsilon_{1})^{2}\left(1-|\langle\psi|\psi_{1}\rangle|^{2}\right)+(\epsilon_{2})^{2}\left(1-|\langle\psi|\psi_{2}\rangle|^{2}\right)+\right.\\
\left.\epsilon_{1}\epsilon_{2}\left(\langle\psi_{2}|\psi_{1}\rangle+\langle\psi_{1}|\psi_{2}\rangle-\langle\psi_{1}|\psi\rangle\langle\psi|\psi_{2}\rangle-\langle\psi_{2}|\psi\rangle\langle\psi|\psi_{1}\rangle\right)\right].
\end{eqnarray*}
 Now we note that $\langle\epsilon_{1}\epsilon_{2}\rangle=\langle\epsilon_{1}\rangle\langle\epsilon_{2}\rangle=0.$
So we are left with: 
\[
\langle(\epsilon_{1})^{2}\rangle\left(1-|\langle\psi|\psi_{1}\rangle|^{2}\right)+\langle(\epsilon_{2})^{2}\rangle\left(1-|\langle\psi|\psi_{2}\rangle|^{2}\right),
\]
 but we already calculated $\langle(\epsilon_{1})^{2}\rangle,\langle(\epsilon_{2})^{2}\rangle:$this
is simply the infidelity of the measurement. Hence in the leading order
of $e$ , the fidelity of the CNOT is: 
\[
\epsilon(2-(|\beta|^{2}+|\delta|^{2}-|\alpha|^{2}-|\gamma|^{2})^{2}-(\alpha^{*}\delta+\delta^{*}\alpha+\beta^{*}\gamma+\gamma^{*}\beta)^{2}),
\]
where $\epsilon$ denotes the measurement infidelity. Note that
by taking for example: $\alpha=\frac{1}{2},\delta=\frac{i}{2},\beta=\frac{1}{2},\gamma=\frac{i}{2},$
the infidelity is just $2\epsilon.$ It is noted in the main text that single body corrections should be applied in the case of wrong measurement outcomes: $\sigma_{X,3}$ correction for the $\sigma_{Z,1}\sigma_{Z,3}$ measurement, $\sigma_{Z,1}\sigma_{Z,3}$ correction for the $\sigma_{X,2}\sigma_{X,3}$ measurement and $\sigma_{X,2}$ correction for the $\sigma_{Z,3}$ measurement. It can be seen that the first and the last correction cancel each other out; thus a correction of $\sigma_{Z,2}$ should only be applied at the end if one of the undesired outcomes was obtained. This accounts for an additional $3\epsilon_{2}$ term in the infidelity.
Therefore the total infidelity of the CNOT as a function of $n,$ the required majority vote, reads: 
\begin{equation}
\left[8n(1+\frac{2e^{2}}{3})+9\right]\epsilon_{2}+{2n-1 \choose n}\left(\frac{2e^{2}}{3}\right)^{n}.
\label{total_infidelity_supp}
\end{equation}

Eq. \ref{total_infidelity_supp} tells us that the problem of decreasing the infidelity of the CNOT is mapped to the problem of reducing the single body errors ($\epsilon_{2}$). This is because if $\epsilon_{2}$ is small enough we can take as many iterations as needed, thereby reducing the amplitude error below the threshold. Given a threshold $T,$ any $\epsilon_{2}$ limits the number of iterations and thus imposes a new threshold for the original CNOT error. This is shown in fig. \ref{new_threshold} in the main part. Note that the  original CNOT error is: $\frac{1}{2}\frac{1}{2e}\overset{e}{\underset{-e}{\int}}\sin^{2}(x)dx=\frac{e^{2}}{6};$ thus the infidelity of our CNOT as a function of the original CNOT error (denoted as $\epsilon$) is: 
\begin{equation}
\left[8n(1+4\epsilon)+9\right]\epsilon_{2}+{2n-1 \choose n}\left(4\epsilon\right)^{n}.
\end{equation}

\end{widetext}

\begin{references}

\bibitem{shor} P. W. Shor,
\newblock Phys. Rev. A 52(4):R2493 (1995). 

\bibitem{knill}
E. Knill, R. Laflamme, and W. Zurek.
\newblock {\it Tec.Rep. LAUR-96-2199 LANL}.

\bibitem{aharonov} D. Aharonov and M. Ben-Or, \href{http://dl.acm.org/citation.cfm?doid=258533.258579}{in Proceedings of the 29th ACM Symposium on Theory of Computing (ACM,New York, 1997), p. 176.}


\bibitem{kitaev}
Kitaev, A. Yu. 
\newblock {Ann. Phys.} 303 2-30 (2003): .

\bibitem{Aliferis}
Aliferis, P., Gottesman, D. and Preskill, J. 2006 Quantum accuracy threshold for concatenated
distance-3 codes. Quantum Inf. Comput. {\bf 6}, 97.


\bibitem{ballance1}
C. J. Ballance, T. P. Harty, N. M. Linke, M. A. Sepiol, D. M. Lucas,  Laser-driven quantum logic gates with precision beyond the fault-tolerant threshold. arXiv:1512.04600.


\bibitem{Du1}
Xing Rong, et. al.,  Experimental fault-tolerant universal quantum gates with solid-state spins under ambient conditions.
Nature Communications 6, 8748 (2015) 

\bibitem{Biercuk1}
M. J. Biercuk. et. al.,  Experimental Uhrig dynamical decoupling using trapped ions. Phys. Rev. A {\bf 79}, 062324 (2009).

\bibitem{Timoney1}
N. Timoney, V. Elman, W. Neuhauser, Chr. Wunderlich. Error-resistant Single Qubit Gates with Trapped Ions.
Physical Review A {\bf 77}, 052334 (2008).

\bibitem{Tan}
T. R. Tan, et. al., Demonstration of a Dressed-State Phase Gate for Trapped Ions. Phys. Rev. Lett. 110, 263002  (2013).

\bibitem{Webster}
S. C. Webster, et. al., Simple Manipulation of a Microwave Dressed-State Ion Qubit. 
Phys. Rev. Lett. {\bf 111}, 140501 (2013).

\bibitem{Timoney2}
N. Timoney, et. al.,  Quantum Gates and Memory using Microwave Dressed States.
Nature {\bf 476}, 185 (2011).

\bibitem{MS}
Anders Sorensen and Klaus Molmer, Quantum Computation with Ions in Thermal Motion. \href{http://journals.aps.org/prl/abstract/10.1103/PhysRevLett.82.1971}
{Phys. Rev. Lett. {\bf 82}, 1971 (1999).}

\bibitem{Wineland2000N}C. A. Sackett, D. Kielpinski, B. E. King, C. Langer, V. Meyer, C. J. Myatt, M. Rowe, Q. A. Turchette, W. M. Itano, D. J. Wineland, and C. Monroe, \href{http://www.nature.com/nature/journal/v404/n6775/full/404256a0.html}{Nature {\bf 404,} 256-259 (2000).}

\bibitem{Blatt2014NPH} J. Benhelm, G. Kirchmair, C. F. Roos, and R. Blatt, \href{http://www.nature.com/nphys/journal/v4/n6/full/nphys961.html}{Nature Physics {\bf 4,} 463 - 466 (2008).}

\bibitem{Monroe2009PRL} K. Kim, M.-S. Chang, R. Islam, S. Korenblit, L.-M. Duan, and C. Monroe \href{http://journals.aps.org/prl/abstract/10.1103/PhysRevLett.103.120502}{Phys. Rev. Lett. {\bf 103,} 120502  (2009).}

\bibitem{Ozeri2014PRA} N. Navon, N. Akerman, S. Kotler, Y. Glickman, and R. Ozeri \href{http://journals.aps.org/pra/abstract/10.1103/PhysRevA.90.010103}{Phys. Rev. A {\bf 90,} 010103(R) (2014).}

\bibitem{Tan2} T. R. Tan,	J. P. Gaebler,	Y. Lin,	Y. Wan,	R. Bowler,	D. Leibfried, and D. J. Wineland, \href{http://www.nature.com/nature/journal/v528/n7582/full/nature16186.html}{Nature {\bf 528,} 380–383 (2015).}

\bibitem{Didi2003Nature} D. Leibfried, B. DeMarco, V. Meyer, D. Lucas, M. Barrett, J. Britton, W. M. Itano, B. Jelenkovic acute, C. Langer, T. Rosenband and D. J. Wineland \href{http://www.nature.com/nature/journal/v422/n6930/full/nature01492.html}{Nature {\bf 422,} 412-415 (2003).}

\bibitem{Lucas2015Nature1} C. J. Ballance,	V. M. Schafer,	J. P. Home,	D. J. Szwer,	S. C. Webster,	D. T. C. Allcock,	N. M. Linke,	T. P. Harty,	D. P. L. Aude Craik,	D. N. Stacey,	A. M. Steane	and D. M. Lucas, \href{http://www.nature.com/nature/journal/v528/n7582/full/nature16184.html}{Nature {\bf 528,} 384–386 (2015).}

\bibitem{Lucas2015arXiv}C. J. Ballance, T. P. Harty, N. M. Linke, M. A. Sepiol, D. M. Lucas, \href{http://arxiv.org/abs/1512.04600}{arXiv:1512.04600 (2015).}

\bibitem{Hayes2012PRA} D. Hayes, S. M. Clark, S. Debnath, D. Hucul, I. V. Inlek, K. W. Lee, Q. Quraishi, and C. Monroe, \href{http://journals.aps.org/prl/abstract/10.1103/PhysRevLett.109.020503}{Phys. Rev. Lett. {\bf 109,} 020503 (2012).}

\bibitem{Green2015PRL}T. J. Green and M. J. Biercuk \href{http://journals.aps.org/prl/abstract/10.1103/PhysRevLett.114.120502}{
Phys. Rev. Lett. {\bf 114,} 120502 (2015).}

\bibitem{Ivanov2015PRA}S. S. Ivanov and N. V. Vitanov, \href{http://journals.aps.org/pra/abstract/10.1103/PhysRevA.92.022333}{Phys. Rev. A {\bf 92,} 022333 (2015).}

\bibitem{gottesman} D. Gottesman and I.L. Chuang, Demonstrating the viability of universal quantum computation using teleportation and single-qubit operations \href{http://www.nature.com/nature/journal/v402/n6760/full/402390a0.html}{Nature {\bf 402} 390-393 (1999)}

\bibitem{Supplemental Material} Supplamental Material





\end{references}
\end{document}